\documentclass{elsart}

\usepackage{graphicx}
\usepackage{amssymb}
\usepackage{amsmath,amssymb}
\usepackage{indentfirst}

\graphicspath{{/home/sklowe/public_html/May_2010/monopole/}}

\begin{document}

\begin{frontmatter}
\title{Search for GUT Monopoles at Super-Kamiokande}
\author{K. Ueno\thanksref{label1}\corauthref{cor1}}, 
\ead{ueno@suketto.icrr.u-tokyo.ac.jp}
\corauth[cor1]{Corresponding Author}
\author{K. Abe\thanksref{label1}\thanksref{label2}},
\author{Y. Hayato\thanksref{label1}\thanksref{label2}},
\author{T. Iida\thanksref{label1}},
\author{K. Iyogi\thanksref{label1}},
\author{J. Kameda\thanksref{label1}\thanksref{label2}},
\author{Y. Koshio\thanksref{label1}\thanksref{label2}},
\author{Y. Kozuma\thanksref{label1}},
\author{M. Miura\thanksref{label1}\thanksref{label2}},
\author{S. Moriyama\thanksref{label1}\thanksref{label2}},
\author{M. Nakahata\thanksref{label1}\thanksref{label2}},
% ...
\\
(The Super-Kamiokande Collaboration)\\

\address[label1]{Kamioka Observatory, Institute for Cosmic Ray Research, The University of Tokyo, Kamioka, Gifu 506-1205, Japan}
\address[label2]{Institute for the Physics and Mathematics of the Universe (IPMU), The University of Tokyo, Kashiwa, Chiba 277-8583, Japan}
% ...

\begin{abstract}
GUT monopoles captured by the Sun's gravitation are expected to catalyze proton decays via the Callan-Rubakov process. In this scenario, protons, which initially decay into pions, will ultimately produce $\nu_{e}$, $\nu_{\mu}$ and $\bar{\nu}_{\mu}$. After undergoing neutrino oscillation, all neutrino species appear when they arrive at the Earth, and can be detected by a 50,000 metric ton water Cherenkov detector, Super-Kamiokande (SK). A search for low energy neutrinos in the electron total energy range from 19 to 55 MeV was carried out with SK and gives a monopole flux limit of F$_{\mathrm{M}}(\sigma_{0}/1\mathrm{mb}) < 6.3 \times 10^{-24} (\beta_{\mathrm{M}}/10^{-3})^{2}$ cm$^{-2}$ s$^{-1}$ sr$^{-1}$ at 90\% C.L., where $\beta_{\mathrm{M}}$ is the monopole velocity in units of the speed of light and $\sigma_{0}$ is the catalysis cross section at $\beta_{\mathrm{M}}=1$.
 The obtained limit is more than eight orders of magnitude more stringent than the current best cosmic-ray supermassive monopole flux limit, F$_{\mathrm{M}} < 1 \times 10^{-15}$ cm$^{-2}$ s$^{-1}$ sr$^{-1}$ for $\beta_{\mathrm{M}} < 10^{-3}$ and also two orders of magnitude lower than the result of the Kamiokande experiment, which used a similar detection method.
\end{abstract}

\end{frontmatter}

\section{Introduction}
Grand Unified Theories (GUTs) predict superheavy magnetic monopoles (GUT monopoles) produced in the very early universe~\cite{t'Hooft,Polyakov}. GUT monopoles are predicted to have appeared as topological defects at the phase transition of vacuum, where the GUT gauge group spontaneously broke to leave the U(1) of electromagnetism. If we take as an example the temperature of the phase transition at which the monopoles were formed to equal 10$^{15}$ GeV and assume the average production rate of about one monopole per horizon at that time~\cite{Kibble}, the density of the monopoles today would exceed the critical density of the universe by more than 14 orders of magnitude. Even if we circumvent this problem by resorting to inflationary universe scenarios~\cite{Guth,Sato}, we cannot avoid a large uncertainty on the monopole flux in the universe since the flux depends on such parameters as monopole mass and the reheating temperature. In fact, due to the wide variety of elementary particle models, several models are compatible with the level of the Parker bound ($\sim$ 10$^{-15}$ cm$^{-2}$ s$^{-1}$ sr$^{-1}$)~\cite{Parker-Turner,Laza,Dar}, and a flux in that range can be relatively easily detected by underground experiments. Arafune \textit{et al.}~\cite{Arafuku} pointed out that copious low energy neutrinos might be emitted when monopoles accumulating inside the Sun catalyze proton decays,
\begin{equation}
p \rightarrow (\rho^{0},\omega,\eta,K^{+},\cdots) + e^{+}(\mathrm{or} \ \ \mu^{+})
\end{equation}
along their paths with cross sections typical of strong interactions via the Callan-Rubakov process~\cite{Rubakov,Callan}. When decay mesons produced by the above process subsequently decay into positive pions, ($(\rho^{0},\omega,\eta,K^{+},\cdots) \rightarrow \pi^{+}$), $\nu_{e}$, $\nu_{\mu}$ and $\bar{\nu}_{\mu}$ are produced\footnote{Under the assumption that monopoles are accumulated in the center of the Sun, pions decaying in flight are negligible due to the high density of hydrogen in the Sun.} by the reactions,
\begin{eqnarray}
\hspace{128pt}
\pi^{+} &\rightarrow& \mu^{+} + \nu_{\mu} \\
\mu^{+} &\rightarrow& e^{+} + \nu_{e} + \bar{\nu}_{\mu}
\end{eqnarray}

\begin{figure}[htbp]
\begin{center}
\includegraphics[width=8cm]{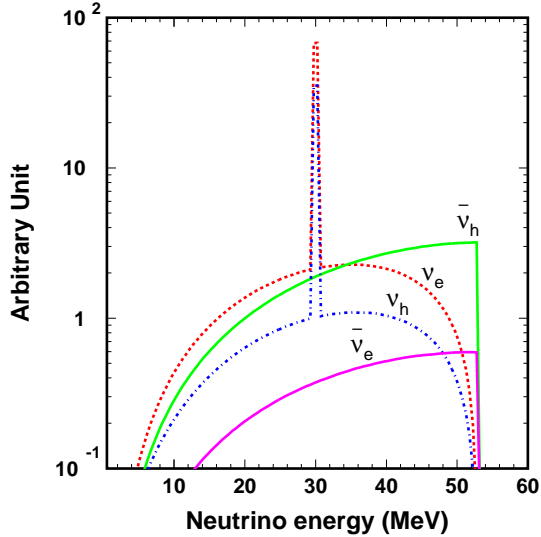}
\caption{Expected neutrino spectra at the SK detector. Neutrino oscillation effects are taken into account. $\mu, \tau$ components are collectively denoted as $h$ since these two flavors of neutrinos have identical cross sections for electron scattering. The spikes at 29.79 MeV originate from $\nu_{\mu}$ produced by the two-body decay of $\pi^{+}$ in Eq. (2), which also oscillates into $\nu_{e}$.}
\label{fig:nue_spec_osc}
\end{center}
\end{figure}

After undergoing neutrino oscillation, all neutrino species are present when they arrive at the Earth (Fig. \ref{fig:nue_spec_osc}), and such neutrinos can be detected by a water Cherenkov detector. Throughout this paper, the neutrinos and antineutrinos are assumed to oscillate with the same parameters of $\sin^{2}{\theta_{12}}$=0.31, $\sin^{2}{\theta_{13}}$=0.02 and $\Delta{m}_{21}^{2}$=$7.6\times10^{-5}$ eV$^{2}$ ~\cite{osc-parameter}. The uncertainties of the mixing angles are taken into account in the systematic errors.

\section{Detector and Expected signal}
We searched for monopole-induced neutrinos using Super-Kamiokande (SK)~\cite{SKdetector}, a large water Cherenkov detector located inside Japan's Kamioka mine. SK is a high-performance neutrino detector consisting of 50,000 metric tons of pure water. Since there are $\gamma$-ray backgrounds near the wall, the fiducial volume for this search, which amounts to 22,500 metric tons, is defined to be more than two meters from the walls of the inner detector (ID).

Monopole-induced neutrinos include all six types, and so for the monopole-induced neutrino search both electron elastic scattering, $\nu_{x}(\bar{\nu}_{x}) + e^{-} \rightarrow \nu_{x}(\bar{\nu}_{x}) + e^{-}$, and inverse beta decay, $\bar{\nu}_{e} + p \rightarrow e^{+} + n$, were assumed to contribute. Figure~\ref{fig:espec_osc} shows the expected spectra of elastic scattering recoil electrons and inverse beta decay positrons in SK. The Mikheyev-Smirnov-Wolfenstein (MSW) effects~\cite{MSW} on oscillation in both the Sun and the Earth were also taken into account. We adopted a simple approximate analytic formula presented in Ref.~\cite{Akhmedov} for the calculation of the MSW effects and assumed the adiabaticity condition is fulfilled, \textit{i.e.}, the density changes so slowly inside the Sun and the Earth that the neutrino mass eigenstates propagate independently. The density profiles of the Sun and the Earth were taken from Refs.~\cite{sun-profile} and \cite{earth-profile}, respectively. 

\section{Data reduction}
In this analysis, we used 2853 live days of data consisting of SK-I (Apr. 1996 - Jul. 2001 : 1497.4 days), SK-II (Oct. 2002 - Oct. 2005 : 793.7 days) and SK-III (Jul. 2006 - Aug. 2008 : 562.0 days). In SK-II the energy resolution was worse because the number of photomultiplier tubes (PMTs) was about half as many as those in the other phases. This difference in detector geometry was taken into account in the calculation of the expected spectra. However, as previously demonstrated by the SK solar neutrino analysis~\cite{solar}, the broad energy spectra typical of monopole-induced neutrinos are hardly affected by the resolution.

\begin{figure}[htbp]
\begin{center}
\includegraphics[width=0.45\linewidth]{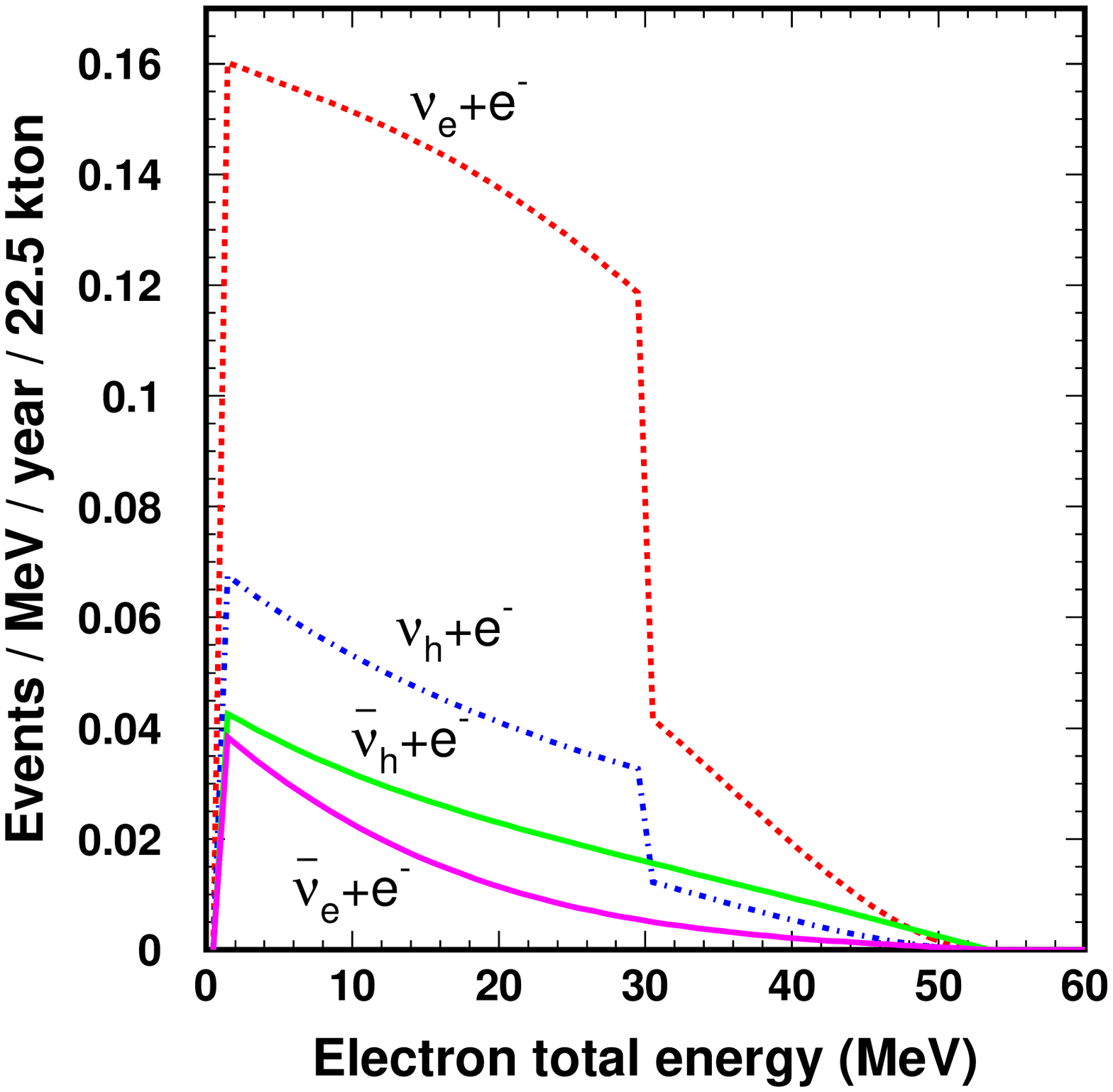}
\includegraphics[width=0.45\linewidth]{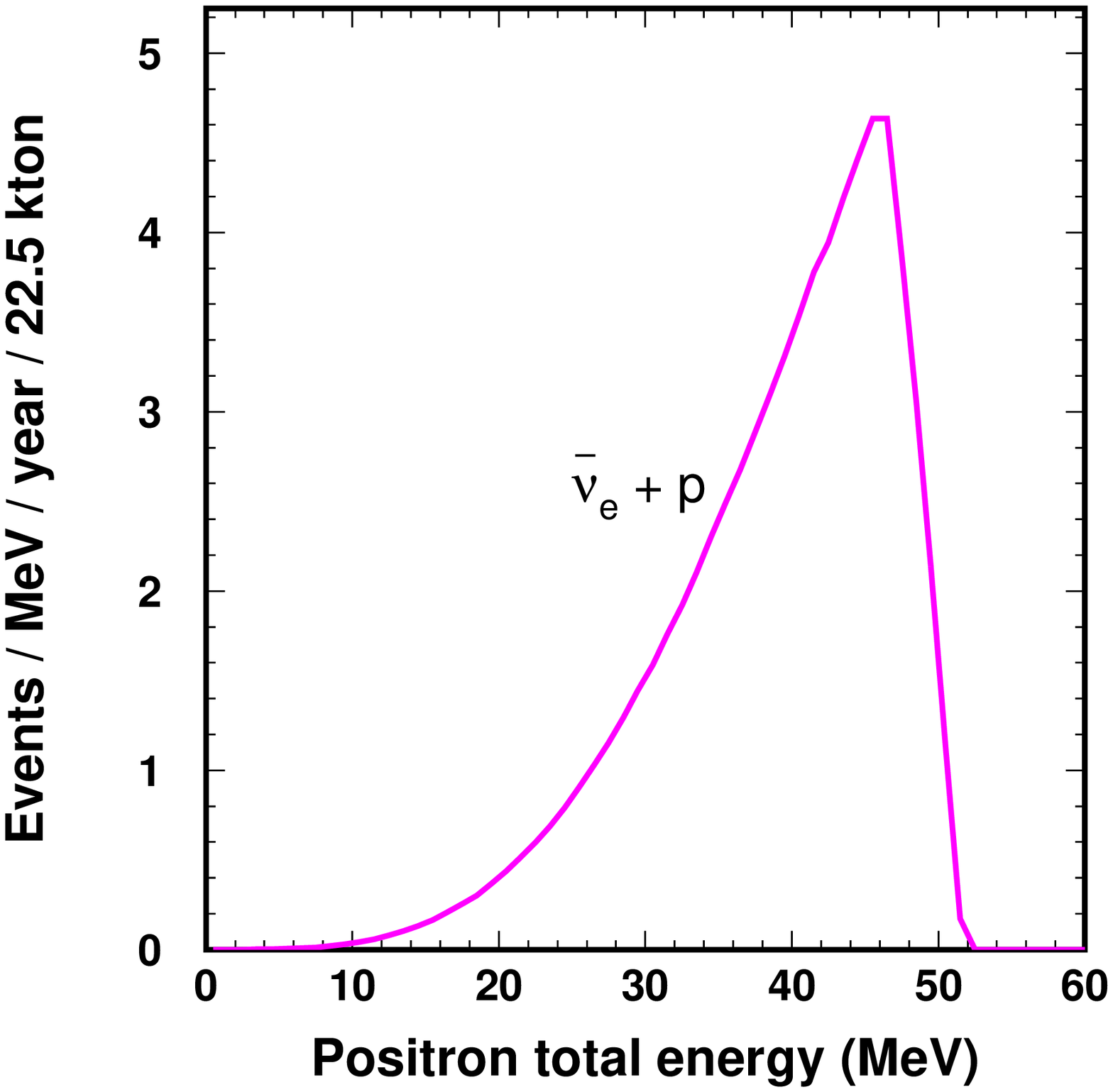}
\caption{Expected energy spectra of elastic scattering recoil electron (left) and inverse beta decay positron (right) in SK assuming neutrino flux of $3.0 \times 10^{2}$ cm$^{-2}$s$^{-1}$ without energy resolution included. The discontinuity of some spectra at 29.79 MeV is caused by the spikes shown in the expected neutrino energy spectra in Fig.~\ref{fig:nue_spec_osc}. $h$ represents $\mu$ and $\tau$ components.}
\label{fig:espec_osc}
\end{center}
\end{figure}

The background events of the monopole-induced neutrino events are mainly caused by the atmospheric neutrinos, the solar neutrinos and muon-induced spallation products. To remove the spallation products and the solar neutrino events, we set the lower energy thresholds to 19 MeV for SK-I and III, and 20 MeV for SK-II in this analysis\footnote{In the SK solar neutrino analysis, we reconstruct the event energy assuming all Cherenkov photons in an event come from a single electron, and we define ``energy'' as the total energy (not the kinetic energy) of the electron. Throughout this paper, we use this definition.}. The upper energy threshold was set to 55 MeV, which was determined by the end-point of the recoil electron (or positron) spectrum (Fig. \ref{fig:espec_osc}).

Besides setting the fiducial volume and energy criteria, some additional background cuts were applied to the data. After the event selection that removes cosmic ray muons and detector noise events, the data sample is subjected to a spallation cut. Cosmic-ray muons can spall oxygen and create unstable nuclei called spallation products ($\mu$ + $^{16}$O $\rightarrow$ $\mu$ + X). This is one of the most abundant backgrounds in the $\lesssim$ 20 MeV region, and as mentioned above, the ability to remove the background events mostly determines the lower threshold for the monopole-induced neutrino search. The spallation background events are reduced by a likelihood method that uses timing, position and photo-electron information of the muons preceding the candidate events.

Then, we applied the Cherenkov opening angle cut to the remaining data sample. The opening angle is estimated by histogramming all the angles uniquely determined by the reconstructed vertex and the possible combinations of three hit PMTs. The histogram is divided into 100 angle bins and the peak is located by finding the successive seven bins with the largest number of entries. The angle corresponding to the midpoint of the seven bins is regarded as the Cherenkov angle of the event. Most of the remaining visible atmospheric (anti-)muon neutrino events are removed by this cut. Electrons with E $>18$ MeV have a Cherenkov angle $\theta_C$ of about 42$^{\circ}$, while visible muons remaining in the signal energy range have momenta less than $\sim$ 250~MeV/c, which corresponds to the maximum $\theta_C$ value of 36$^{\circ}$. Thus, taking into account the finite resolution of the Cherenkov angle measurements, events with $\theta_C$ $<$ 38$^{\circ}$ were removed. Also we removed events with $\theta_C$ $>$ 50$^{\circ}$. Those events do not have a clear Cherenkov ring pattern and originate from multiple $\gamma$-rays emitted by excited nuclei created in neutral current (NC) interactions of atmospheric neutrinos.

Some events originating from the outside of the fiducial volume are reconstructed within the fiducial volume of SK. These events are $\gamma$-rays from the materials of the detector structure and the surrounding rock. To remove such events, we cut on the event's distance to the ID wall projected backwards from its vertex position along its reconstructed direction. Events with this distance less than 300 cm were removed ($\gamma$-ray cut~\cite{solar})\footnote{Additionally, energy-dependent cut criteria were also applied.}.

We removed charged $\pi$ background events based on the sharpness of the Cherenkov ring pattern. As a charged $\pi$ interacts with or is absorbed by a nucleus and is less influenced by multiple Coulomb scattering than electrons, it generates a sharp Cherenkov ring pattern. Events having two or more Cherenkov rings were also removed.

The remaining backgrounds include low energy muons produced by atmospheric $\nu_{\mu}$ charged current (CC) interactions and their decay electrons. Those events are removed using time and spatial correlations; any event pairs having a time difference and a spatial distance less than 50 $\mu$s and 500 cm, respectively, were removed. In case the candidate event is the second one in a pair, we removed event pairs with the first events having total photo-electrons (p.e.) more than 1000 p.e. for SK-I and III, and 500 p.e. for SK-II, which correspond to $90 \sim 100$ MeV electron equivalent energy. Although this selection (decay electron cut) removes some fraction of the atmospheric neutrino events, there still remain decay electrons which are not accompanied by nuclear de-excitation $\gamma$-rays, and whose parent muons are invisible because their energies are below Cherenkov threshold. Those background events are distinguished from the signal events using angular correlation with the Sun direction, which will be described later.

\section{Analysis and Results}
The numbers of events remaining after each selection step are summarized in Table~\ref{tab:reduction}. We found 317 candidate events in the final data sample. Three types of background events are considered.

\begin{table}[htbp]
\caption{Summary of the number of remaining events and the cut signal efficiency for electron scattering and inverse beta decay after each selection.}
\label{tab:reduction}
\begin{center}
{ \scriptsize
\begin{tabular}{cccc}
\hline\hline
Cut			    & Data  & Detection eff.&    Detection eff.  \\ 
 			    &       & ($\nu$ + e)   &  ($\bar{\nu}_{e}$ + p) \\
\hline
Energy range cut,           &      &      &      \\
Fiducial volume cut,        & 2108 & 0.99 & 0.99 \\
Noise event rejection       &  	   &      &      \\
Spallation event cut        & 1774 & 0.91 & 0.98 \\
Cherenkov opening angle cut & 759  & 0.86 & 0.93 \\
$\gamma$-ray cut	    & 738  & 0.83 & 0.90 \\
Pion-like event cut         & 705  & 0.81 & 0.88 \\
Multi-ring cut              & 686  & 0.81 & 0.87 \\
Decay electron cut          & 317  & 0.81 & 0.87 \\
\hline\hline
\end{tabular}
}
\end{center}
\end{table}

The first dominant source of background events are decay electrons from invisible muons induced by the atmospheric (anti-)muon neutrinos. The second such source are electrons generated by the atmospheric (anti-)electron neutrinos, while the third are the multiple nuclear de-excitation gamma-rays produced by NC interactions of the atmospheric neutrinos.

The remaining spallation events are estimated to be $4.7 \pm 4.3$ events in SK-I/III and $0.2 \pm 1.7$ events in SK-II using the likelihood distribution.

The number of remaining solar neutrino events is estimated to be one.

\begin{figure}[htbp]
\begin{center}
\includegraphics[width=8cm]{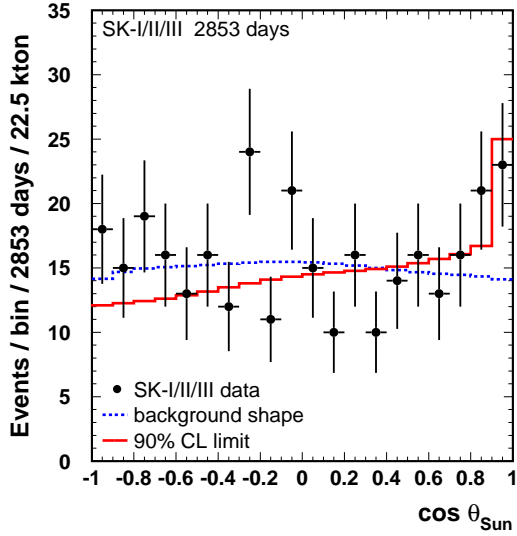}
\caption{Angular distributions with respect to the expected neutrino direction from the Sun. The points with error bars show the data events. The solid and dashed histograms represent $\mathrm{I_{0}}$ = $\mathrm{I_{90}}$ and $\mathrm{I_{0}}$ = 0, respectively, which are described in the text.}
\label{fig:cossun_obs}
\end{center}
\end{figure}

\begin{figure}[htbp]
\begin{center}
\includegraphics[width=0.45\linewidth]{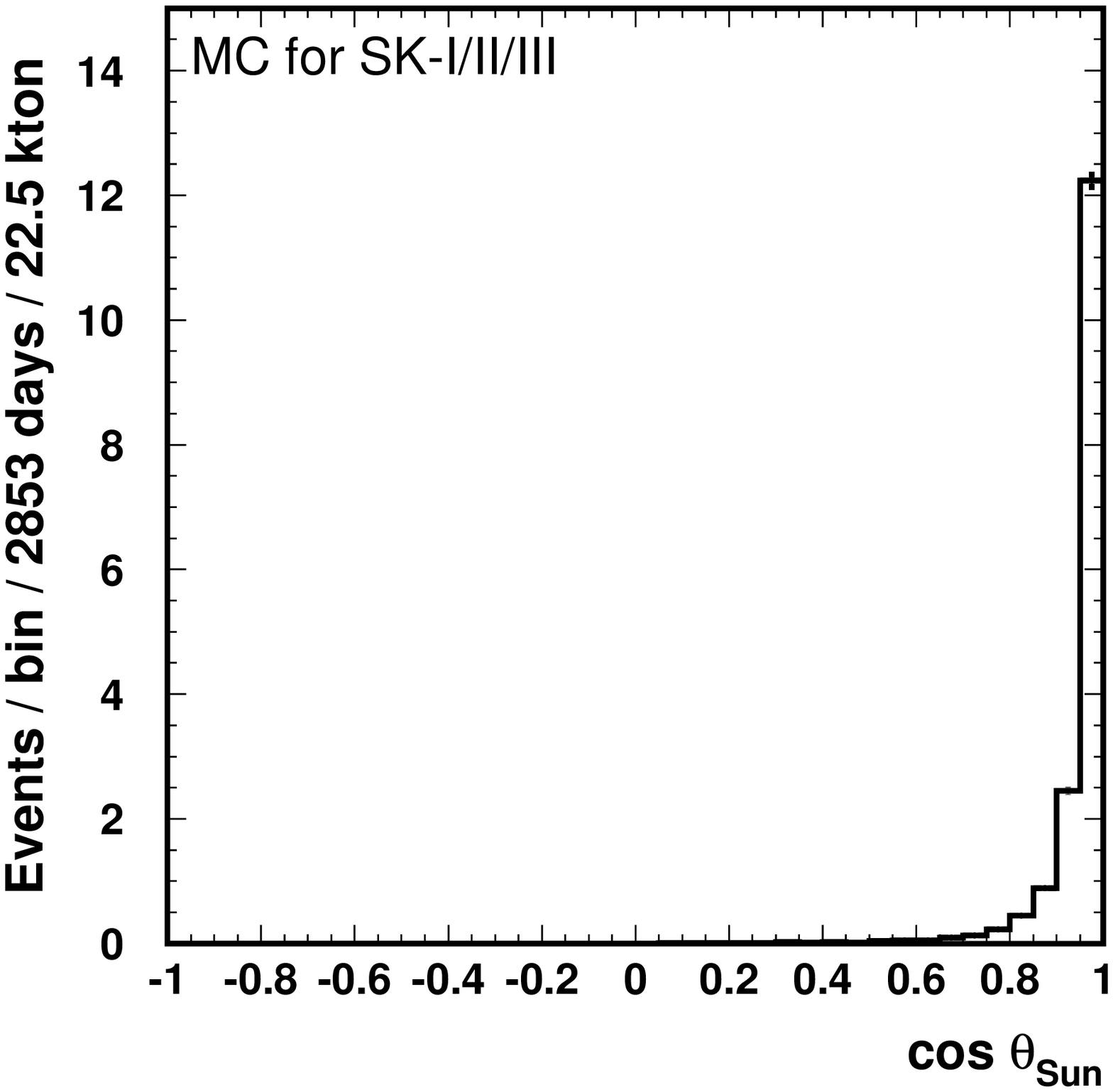}
\includegraphics[width=0.45\linewidth]{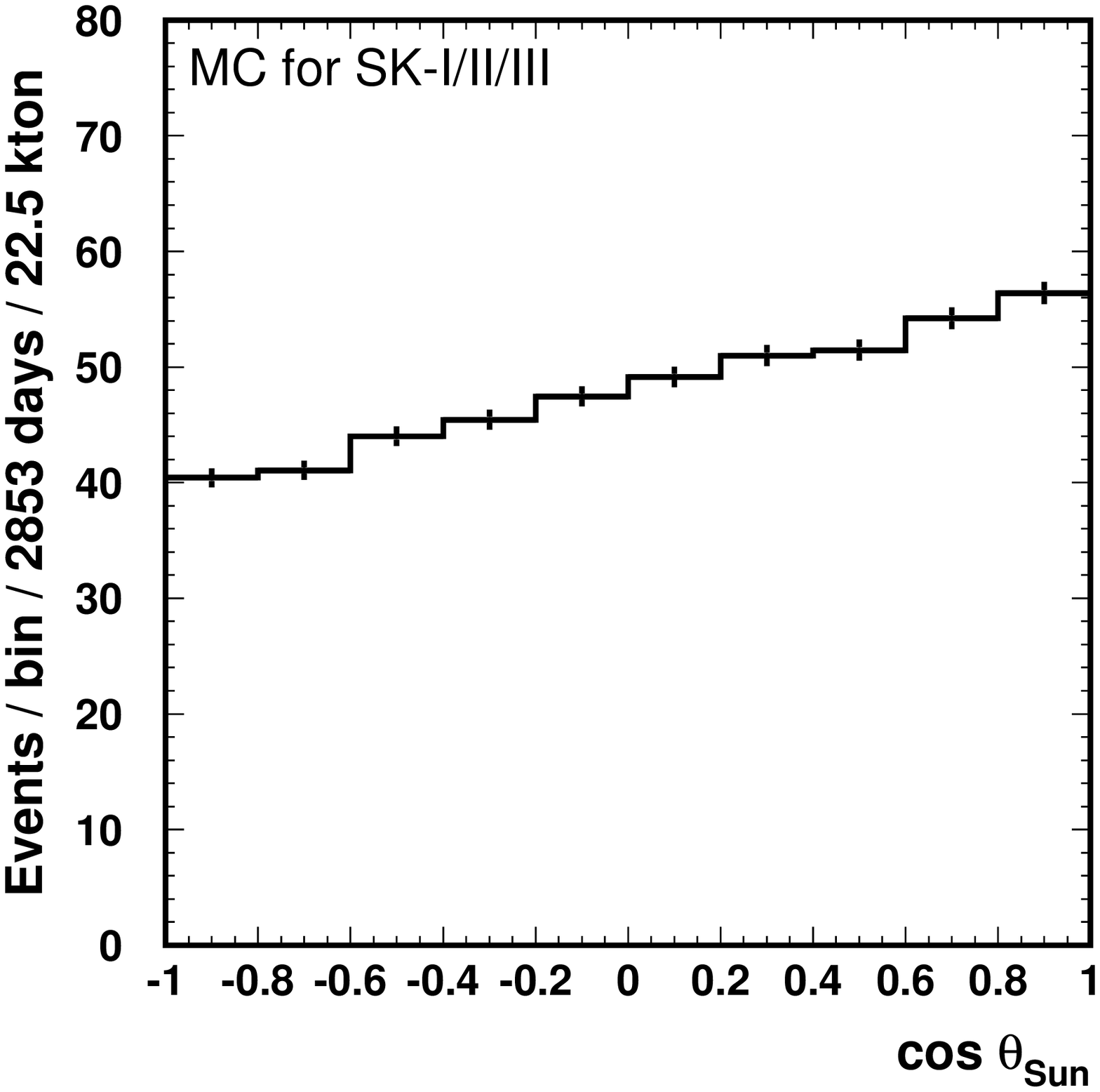}
\caption{Expected angular distributions of elastic scattering recoil electrons (left) and inverse beta decay positrons (right) assuming a neutrino flux of $3.0 \times 10^{2}$ cm$^{-2}$s$^{-1}$ and the detection efficiency of the cuts.}
\label{fig:cossun_mc}
\end{center}
\end{figure}

To extract signal events, an angular distribution with respect to the Sun direction is used. Figure~\ref{fig:cossun_obs} shows the $\cos{\theta_{\mathrm{Sun}}}$ distribution of the 317 candidates, where $\theta_{\mathrm{Sun}}$ is the angle of the candidate event direction with respect to the expected neutrino direction calculated from the position of the Sun at the event time. To fit the distribution, we use a $\chi^{2}$ defined in Eq.(4):
\begin{equation}
\chi^{2}(\mathrm{I_{0}},\alpha) \equiv \sum_{i=1}^{20} \frac{\{N_{\mathrm{obs},i} - N_{\nu e,i}(\mathrm{I_{0}}) - N_{\nu p,i}(\mathrm{I_{0}}) - \alpha N_{\mathrm{bkg},i} \}^{2}}{\sigma_{\mathrm{stat},i}^{2} + \sigma_{\mathrm{sys},i}^{2}},
\end{equation}
where we use $N_{\mathrm{obs},i}$ as the observed number of events in the $i$-th angular bin. $N_{\nu e,i}(\mathrm{I_{0}})$ and $N_{\nu p,i}(\mathrm{I_{0}})$ are the expected signal events from electron scattering and inverse beta decay in the $i$-th bin, where $\mathrm{I_{0}}$ is the total flux of all flavors of neutrinos in units of cm$^{-2}$s$^{-1}$ (Fig. \ref{fig:cossun_mc}). The theoretical calculation of the angular distribution of the inverse beta decay is from Ref.~\cite{Vissani}. The systematic errors used in the $\chi^{2}$ function ($\sigma_{\mathrm{sys},i}= {\rm const.}$) are summarized in Table~\ref{tab:sys_err}.

\begin{table}[htbp]
\caption{Systematic errors (\%) used in the $\chi^{2}$ function. The systematic error on the cross section of inverse beta decay is from Ref.~\cite{Strumia-Vissani}.}
\label{tab:sys_err}
\begin{center}
{ \scriptsize
\begin{tabular}{cccc}
\hline\hline
Error source	     & SK-I & SK-II & SK-III \\ 
\hline
Cross section        & 0.4\% & 0.4\% & 0.4\% \\ 
Neutrino propagation & 5.0\% & 5.0\% & 5.0\% \\ 
Reduction efficiency & 2.3\% & 3.8\% & 2.1\% \\ 
Fiducial volume      & 1.3\% & 1.1\% & 1.0\% \\ 
Livetime             & 0.1\% & 0.1\% & 0.1\% \\ 
\hline
Total                & 5.7\% & 6.4\% & 5.6\% \\
\hline\hline
\end{tabular}
}
\end{center}
\end{table}

$N_{\mathrm{bkg},i}$ is the area-normalized background events for the $i$-th bin. We determine the $N_{\mathrm{bkg},i}$ values using the final data sample by applying an iterative procedure (described in Appendix) to remove a possible bias caused by the solar-correlated events. The factor $\alpha$ representing the background normalization is determined so that the $\chi^{2}$ value is minimized for each $\mathrm{I_{0}}$ value. In the fit, we restrict $\alpha \ge 0$. As a result of the fit, we obtain $\mathrm{I_{0}}$ = 102 $\pm$ 64 $\mathrm{cm^{-2}s^{-1}}$, corresponding to a 1.6 $\sigma$ excess.

We determine the 90$\%$ C.L. upper limit flux $\mathrm{I_{90}}$ by using the following
equation: 
\begin{equation}
\frac{\displaystyle \int_{0}^{\mathrm{I_{90}}} \exp\left(-\chi^{2}(\mathrm{I}_{0})/2\right) \mathrm{dI}_{0}}{\displaystyle \int_{0}^{\infty} \exp\left(-\chi^{2}(\mathrm{I}_{0})/2\right) \mathrm{dI}_{0}} = 0.9
\end{equation}

\begin{figure}[htbp]
\begin{center}
\includegraphics[width=7cm]{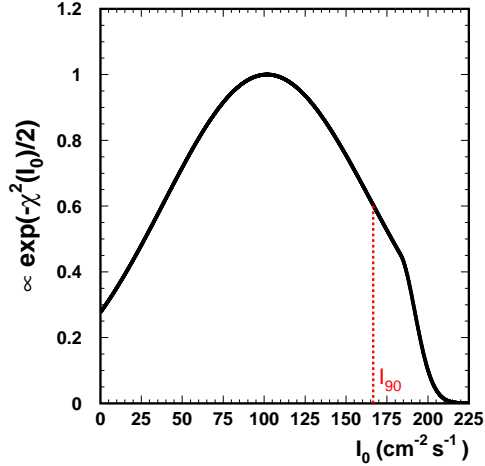}
\caption{Likelihood function ($\propto \exp\left(-\chi^{2}(\mathrm{I}_{0})/2\right)$) as a function of monopole-induced neutrino flux ($\mathrm{I_{0}}$). The dashed line represents $\mathrm{I_{0} = I_{90}}$. The deviation from the Gaussian shape beyond $\mathrm{I}_{0}=183.4\ \mathrm{cm^{-2}s^{-1}}$ is due to the restriction of $\alpha \ge 0$.}
\label{fig:chi2}
\end{center}
\end{figure}

Figure~\ref{fig:chi2} shows the $\exp\left(-\chi^{2}(\mathrm{I}_{0})/2\right)$ distribution as a function of $\mathrm{I_{0}}$, where the normalization is scaled so that the maximum value becomes one. We obtain $\mathrm{I_{90}}$ = 166.6 $\mathrm{cm^{-2}s^{-1}}$. Figure~\ref{fig:cossun_obs} shows the $\cos{\theta_{\mathrm{Sun}}}$ distribution superimposed with curves representing $\mathrm{I_{0}}$ = $\mathrm{I_{90}}$ and $\mathrm{I_{0}}$ = 0.  We obtain a $\chi^{2}(\mathrm{I}_{0}=0)$ value of 20.02 compared to 19 degrees of freedom, which corresponds to a probability of 39.3$\%$ and is therefore consistent with a hypothesis of null signal in the final data sample.

The monopole-catalyzed proton decay rate $f_{p}$ in the Sun is given by,
\begin{equation}
f_{p} = \int n_{\mathrm{M}} v_{\mathrm{rel}} \sigma \rho_{\mathrm{p}} \mathrm{N_{A}} d^{3}x \ \mathrm{decays/s},
\end{equation}
where $n_{\mathrm{M}}$ is the monopole number density, $v_{\mathrm{rel}}=\beta_{\mathrm{rel}}c$ the relative velocity between the monopole and the proton, $\sigma=\sigma_{0}/\beta_{\mathrm{rel}}^{2}$ ($\sigma_{0}\sim$ 1 mb) the catalysis cross section~\cite{Rubakov,Callan,Arafuku2}, $\rho_{\mathrm{p}}$ the proton mass density, and $\mathrm{N_{A}}$ Avogadro's number ($6.0 \times 10^{23}$). The space integral is taken over the interior of the Sun. Helium gives a negligible contribution~\cite{Arafuku2}. If we assume the monopoles are accumulated in the center of the Sun where $\rho_{\mathrm{p}}=50$ $\mathrm{g/cm^{3}}$ and $\beta_{\mathrm{rel}}=1.7\times 10^{-3}$\footnote{We take $v_{\mathrm{rel}}$ as the thermal velocity of proton in the center of the Sun, $\beta_{\mathrm{rel}}c = (2k_{\mathrm{B}}T_{\mathrm{c}}/m_{p})^{1/2} \simeq 1.7 \times 10^{-3}$ m/s, where $k_{\mathrm{B}}$ is the Boltzmann constant, $T_{\mathrm{c}}=1.5\times 10^{7}$ K is the temperature at the center of the Sun, and $m_{p}=0.938$ GeV is the proton mass.}, we obtain $f_{p}=1.7\times 10^{6} (\sigma_{0}/1\ \mathrm{mb}) \mathrm{N_{M}}$, where $\mathrm{N_{M}}$ is the integrated number of monopoles in the region. The rate $f_{p}$ is given by $f_{p}=4\pi d^{2} \mathrm{I}_{0}/3 /f_{\nu_{e}}$, where $d = 1.5 \times 10^{13}$ cm is the distance between the Sun and the Earth. The factor 3 takes into account the fact that three neutrinos are emitted per proton decay. $f_{\nu_{e}}=f_{\pi^{+}}(1-a_{\pi^{+}})$ is the fraction of $\nu_{e}$ produced in a proton decay, where $f_{\pi^{+}}$ is the branching fraction of a proton decay into $\pi^{+}$ + anything, and is about 0.5 for some GUT models. The factor $a_{\pi^{+}}=0.2$ is the absorption probability of $\pi^{+}$ at the center of the Sun~\cite{Arafuku}. Using the $\mathrm{I}_{90}$ value, we obtain an upper limit on $\mathrm{N_{M}}$ at the 90\% C.L.: 
\begin{equation}
\mathrm{N_{M}}\left(\frac{\sigma_{0}}{1\ \mathrm{mb}}\right) \left(\frac{f_{\pi^{+}}}{0.5}\right) < 7.1 \times 10^{17}\ (90\%\mathrm{C.L.})
\label{eqn:nm-limit}
\end{equation}
The monopole flux is calculated from the following equation by assuming monopole-antimonopole annihilation is negligible: 
\begin{equation}
\mathrm{N_{M}} = \pi R_{\odot}^{2} \left(1 + \left(\frac{\beta_{\mathrm{esc}}}{\beta_{\mathrm{M}}}\right)^{2}\right) \epsilon(\mathrm{M_{M}},\beta_{\mathrm{M}}) 4 \pi F_{\mathrm{M}} t_{\odot},
\label{eqn:nm-fm}
\end{equation}
where $\beta_{\mathrm{esc}} = 2 \times 10^{-3}$ and $\beta_{\mathrm{M}}$ are the escape velocity and the monopole velocity in units of the speed of light, respectively, $R_{\odot}$ is the solar radius ($7.0 \times 10^{10}$ cm) and $t_{\odot}$ the elapsed time after the birth of the Sun ($4.6 \times 10^{9}$ yr). $\epsilon(\mathrm{M_{M}},\beta_{\mathrm{M}})$ is the efficiency with which the Sun captures monopoles that strike its surface, which is dependent on the monopole mass $\mathrm{M_{M}}$ and velocity $\beta_{\mathrm{M}}$. We estimate $\epsilon(\mathrm{M_{M}},\beta_{\mathrm{M}})$ based on the calculation of the energy loss of a monopole inside the Sun from Refs.~\cite{Hamilton, Meyer}. From (\ref{eqn:nm-limit}) and (\ref{eqn:nm-fm}), we obtain for $\beta_{\mathrm{M}} < 10^{-3}$
\begin{equation}
F_{\mathrm{M}}\left(\frac{\sigma_{0}}{1\ \mathrm{mb}}\right)\left(\frac{f_{\pi^{+}}}{0.5}\right) < 6.3 \times 10^{-24} \left( \frac{\beta_{\mathrm{M}}}{10^{-3}} \right)^{2}\ \mathrm{cm^{-2}s^{-1}sr^{-1}} \ (90\%\ \mathrm{C.L.})\ \ \ 
\end{equation}
This limit is only valid for $\mathrm{M_{M}}\lesssim 10^{17}$ GeV since $\epsilon(\mathrm{M_{M}},\beta_{\mathrm{M}})$ decreases as the monopole mass increases and becomes significantly less than 1 for higher monopole masses.


\begin{figure}[htbp]
\begin{center}
\includegraphics[width=8cm]{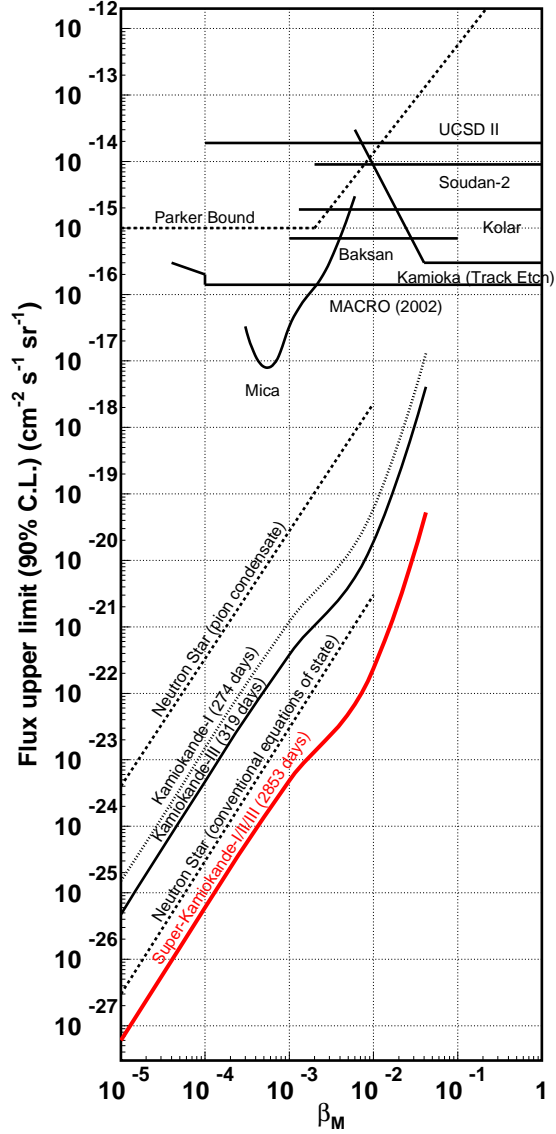}
\caption{90$\%$ C.L. upper limits on the monopole flux as a function of monopole velocity, $\beta_{\mathrm{M}}$. Each flux limit is referenced in the text. Catalysis cross section ($\sigma_{0}$=1mb) and monopole mass ($\mathrm{M_{M}}=10^{16}$ GeV) are assumed for indirect searches.}
\label{fig:monopole-flux-limit}
\end{center}
\end{figure}

Our result is presented in Fig.~\ref{fig:monopole-flux-limit}. Also shown are monopole flux limits obtained by various recent direct detection experiments: UCSD-II \cite{UCSD}, Soudan-2 \cite{Soudan}, Baksan \cite{Baksan}, Kolar \cite{Kolar}, Kamioka track etch \cite{Orito}, Mica \cite{Mica} and the MACRO final result \cite{MACRO}, as well as ones from indirect detection experiments: the Parker bound \cite{Parker-Turner}, limits from neutron star observations \cite{Kolb-Turner}, and the Kamiokande experimental results \cite{Kajita,Sakai}. As seen in the figure, our new limit is many orders of magnitude more stringent than those of the direct experiments and is also two orders of magnitude lower than the result of the Kamiokande experiment~\cite{Kajita,Sakai}, which used a similar detection method. This is mainly because SK is a larger detector and the background events are subtracted effectively using angular distributions. The present limit is lower than the one obtained from the X-ray excess of old neutron stars,
\begin{equation}
F_{\mathrm{M}}\left(\frac{\sigma_{0}}{1\mathrm{mb}}\right) < 3 r \times 10^{-23}\ \mathrm{cm^{-2}s^{-1}sr^{-1}},
\end{equation}
where $\beta_{\mathrm{M}}$ is $10^{-3}$ and $r$ is the ratio of the total luminosity to the photon luminosity of old neutron stars. The value of $r$ varies from 1 to $10^{4}$, where $r \sim 1$ corresponds to the conventional neutron star equations of state while large values of $r$ correspond to the possibility of a pion condensate\footnote{If there is a pion condensate in a neutron star, its total luminosity is dominated by neutrinos generated in the following reactions: $\pi^{-} + n \rightarrow n + \ell^{-} + \bar{\nu}_{\ell}$ and $n + \ell^{-} \rightarrow n + \pi^{-} + \nu_{\ell}$, where $\ell = e$ or $\mu$.} or quark matter core~\cite{Kolb-Turner}. $\beta_{\mathrm{rel}}$ in old neutron stars is of order of 0.1 $\sim$ 0.3. Therefore, our limit is better than the neutron star limit even in the most stringent case. Finally, we emphasize that this result also provides constraints on the parameter space of inflation models as suggested in ~\cite{Laza,Dar}.

\section{Conclusion}
A search for monopole-induced neutrinos from the Sun in the electron total energy range from 19 to 55 MeV was carried out using a data sample of 2853 days' livetime and 22,500 metric tons fiducial volume. We found no signal events and set the 90\% confidence level limit on the monopole flux: F$_{\mathrm{M}}(\sigma_{0}/1\mathrm{mb}) < 6.3 \times 10^{-24} (\beta_{\mathrm{M}}/10^{-3})^{2}$ cm$^{-2}$ s$^{-1}$ sr$^{-1}$. Our result provides the world's most stringent upper limit for $\beta_{\mathrm{M}} < 10^{-2}$.

\section{Acknowledgments}
We gratefully acknowledge the cooperation of Kamioka Mining and Smelting Company.
The Super-Kamiokande experiment was built and has been operated with funding from the Japanese Ministry of Education, Culture, Sports, Science and Technology, the United States Department of Energy, the U.S. National Science Foundation, and the National Natural Science Foundation of China. This work was supported by a Grant-in-Aid for Scientific Research. We thank F. Vissani for providing to us the angular distribution of the inverse beta decay positrons.

\appendix 
\def\thesection{Appendix \Alph{section}}
\section{Estimation of \{$N_{\mathrm{bkg},i}$\}}
Although the background shape in the angular distribution with respect to the direction of the Sun is expected to be almost isotropic, we tried rejecting directional biases including those caused by solar-correlated events remaining in the final data sample, which will distort the $\cos{\theta_{\mathrm{Sun}}}$ distribution.

We define N$_{\mathrm{fin}}$ as the number of the final candidates in each SK phase and prepare event direction vectors, $\boldsymbol{d}_{i}$, and solar direction vectors, $\boldsymbol{s}_{j}$ for the final sample ($i,j=1,\cdots,\mathrm{N}_{\mathrm{fin}}$). Then, the angle with respect to the direction of the Sun of the $i$-th event is expressed as $\cos{\theta_{\mathrm{Sun},i}} = \boldsymbol{d}_{i} \cdot \boldsymbol{s}_{i}$. 

First we fill the $\mathrm{N_{fin}}(\mathrm{N_{fin}}-1)/2$ possible combinations of $\boldsymbol{d}_{i} \cdot \boldsymbol{s}_{j}$ ($i \neq j$) into a histogram $h_{1}$, and fit $h_{1}$ with a polynomial $f_{1}(x), x \in [-1,1]$. Also, we fit the actual true $\cos{\theta_{\mathrm{Sun}}}$ distribution with another polynomial $c(x)$.

Then we fill the $\mathrm{N_{fin}}(\mathrm{N_{fin}}-1)/2$ combinations of $\boldsymbol{d}_{i} \cdot \boldsymbol{s}_{j}$ ($i \neq j$) into another histogram $h_{2}$ with weights of $w_{1}(\boldsymbol{d}_{i} \cdot \boldsymbol{s}_{j}) \equiv f_{1}(\boldsymbol{d}_{i} \cdot \boldsymbol{s}_{i})/c(\boldsymbol{d}_{i} \cdot \boldsymbol{s}_{i})$.

Again we fit $h_{2}$ with another polynomial $f_{2}(x)$.

Usually, $f_{i}(x),w_{i}$ and $h_{i}$ sufficiently converge after several iterations of the above procedure. In this way, we eliminate the bias from solar-correlated events in the limit that $f_{i}(x)$ describes the histogrammed distribution.

\end{document}